# Post-CMOS Compatible Aluminum Scandium Nitride/2D Channel Ferroelectric Field-Effect-Transistor

Xiwen Liu,[a] Dixiong Wang,[a] Jeffrey Zheng,[b] Pariasadat Musavigharavi,[a,b] Jinshui Miao,[a] Eric A. Stach,[b,c] Roy H. Olsson III,[a] Deep Jariwala[a]*

[a] Electrical and Systems Engineering, University of Pennsylvania, Philadelphia, PA, USA

[b] Material Science and Engineering, University of Pennsylvania, Philadelphia, PA, USA

[c] Laboratory for Research on the Structure of Matter, University of Pennsylvania, Philadelphia, PA, USA

*Corresponding author: dmj@seas.upenn.edu

In 1963, Moll and Tarui [1] suggested that the field-effect conductance of a semiconductor could be controlled by the remanent polarization of a ferroelectric (FE) material to create a ferroelectric field-effect transistor (FE-FET). However, subsequent efforts to produce a practical, compact FE-FET have been plagued by low retention and incompatibility with Complementary Metal Oxide Semiconductor (CMOS) process integration. These difficulties led to the development of trapped-charge based memory devices (also called floating gate or flash memory), and these are now the mainstream non-volatile memory (NVM) technology [2]. Over the past two decades, advances in oxide FE materials have rejuvenated the field of ferroelectrics and made FE random access memories (FE-RAM) a commercial reality [3-5]. Despite these advances, commercial FE-RAM based on lead zirconium titanate (PZT) has stalled at the 130 nm due to process challenges [6].The recent discovery of scandium doped aluminum nitride (AlScN) as a CMOS compatible ferroelectric [7] presents new opportunities for direct memory integration with logic transistors due to the low temperature of AlScN deposition (approx. 350 °C). This temperature is compatible with CMOS back end of line processes. Here, we present a FE-FET device composed of an AlScN FE dielectric layer integrated with a channel layer of a van der Waals two-dimensional (2D) semiconductor, $MoS_2$. Our devices show an ON/OFF ratio ~ $10^6$, concurrent with a normalized memory window of 0.3 V/nm. The devices also demonstrate stable, two-state memory retention for up to $10^4$ seconds. Our simulations and experimental results suggest that the combination of AlScN and 2D semiconductors is nearly ideal for low power FE-FET memory. Furthermore, the low processing temperatures required for device fabrication allow extreme scaling because the devices are compatible with back-end-of-the-line (BEOL) Si CMOS. These results demonstrate a new approach in embedded memory and in-memory computing, and could even lead to effective neuromorphic computing architectures.

## Introduction

The slow-down in Moore's Law [8] and the emergence of the memory bottleneck in utilizing Big-Data [9] has created an urgent need for low-power, highly scalable memory devices. These needs require the development of computing hardware architectures different from the standard von Neumann architecture and require tight integration with on-chip memory devices. [10-11]. New applications such as the Internet of Things (IoT) and artificial intelligence (AI) algorithms - applications that either generate or consume vast amounts of data [12-13] - create a strong demand for high-density non-volatile memory (NVM). Among the various emerging NVM technologies, ferroelectric (FE) based NVM devices are the most compelling due to their simple device structure, higher access speed, high endurance, and extremely low write energy [14-18]. Although ferroelectric random-access memory (FE-RAM) is an extant commercial technology, the device architecture requires an FE capacitor to be connected in series with a transistor. This results in the formation of a one transistor-one capacitor (1T-1C) structure that is very similar to conventional dynamic RAM (DRAM). However, the readout of a FeRAM cell is destructive: the process of reading overwrites the FE capacitor by switching its polarity to extract the read current signal. This is not the case with a FE-FET, a compact and low-power memory technology with exceptional promise for dense scaling. Several persistent challenges have prevented the creation of scalable and durable FE-FETs. They include:

1. the lack of a ferroelectric material with sufficiently large coercive field and remanent polarization [15-18]
2. the incompatibility of viable ferroelectric dielectrics with standard CMOS processing [17]
3. poor retention due to both the large depolarization fields caused by the potential drop across the interfacial dielectric and semiconductor band bending. [18]

Here, we demonstrate a high-performance FE-FET that integrates an atomically thin, two-dimensional (2D) molybdenum disulfide ($MoS_2$) channel on top of an AlScN dielectric. The devices achieve an ON/OFF ratio ~$10^6$ between the two memory states of "0" and "1". The AlScN dielectric is deposited onto a Si substrate at temperatures below 350 ℃, making the process compatible with back-end-of-the-line (BEOL) CMOS integration. We build upon recent reports showing that the remanent polarization of AlScN can be very high (80-115 μC/cm2) when Sc concentrations exceed 27% [7]. The key advantage of the high remanent polarization is that the instabilities induced by both charge trapping and leakage currents through the ferroelectric insulator do not significantly affect the FE-FET device performance. Also, the large coercive fields exhibited by AlScN (2-4.5 MV/cm) effectively screen the depolarization fields, which helps achieve long retention times. The superior ferroelectric properties and CMOS compatibility of AlScN and the low-temperature deposition/transfer processing of van der Waals semiconductors suggest that this approach could lead to a new generation of scalable, high-performance, and low-power memory devices compatible with Si CMOS processors.

**Structure of $MoS_2$ FE-FETs with ferroelectric AlScN**

The AlScN/$MoS_2$ FE-FETs are bottom gated transistors with 100 nm thick $Al_{0.71}Sc_{0.29}N$ grown by sputter deposition on 100 nm thick Pt template (Figure 1 a) that was deposited on a heavily doped Si (100) wafer. Pt templates the AlScN with registry along [0001] axis direction, yielding a highly textured FE dielectric. The Pt acts as the gate electrode, and titanium/gold are deposited by electron-beam evaporation to form the source-drain contacts. A cross-sectional bright-field transmission electron microscopy (TEM) image of a representative AlScN/$MoS_2$ FE-FET is shown in Fig. 1b, and a detailed energy-dispersive X-ray spectrometry (EDS) elemental map is presented in Fig. S3. The EDS analysis confirms the presence and uniform distribution of Al, Sc, N, Mo and S. The corresponding selected area electron diffraction (SAED) image pattern shows that the ferroelectric AlScN is highly crystalline and textured along the [0001] growth direction, evident in the arc of the 0002 reflection (Figure 1c). A high-resolution phase-contrast TEM image of the AlScN/$MoS_2$ interface is shown in Fig. 1d, and the few-layer $MoS_2$ channel layer is clearly visible. A thin oxide layer is also observed on the top surface of AlScN; it is present in all samples used in this study.

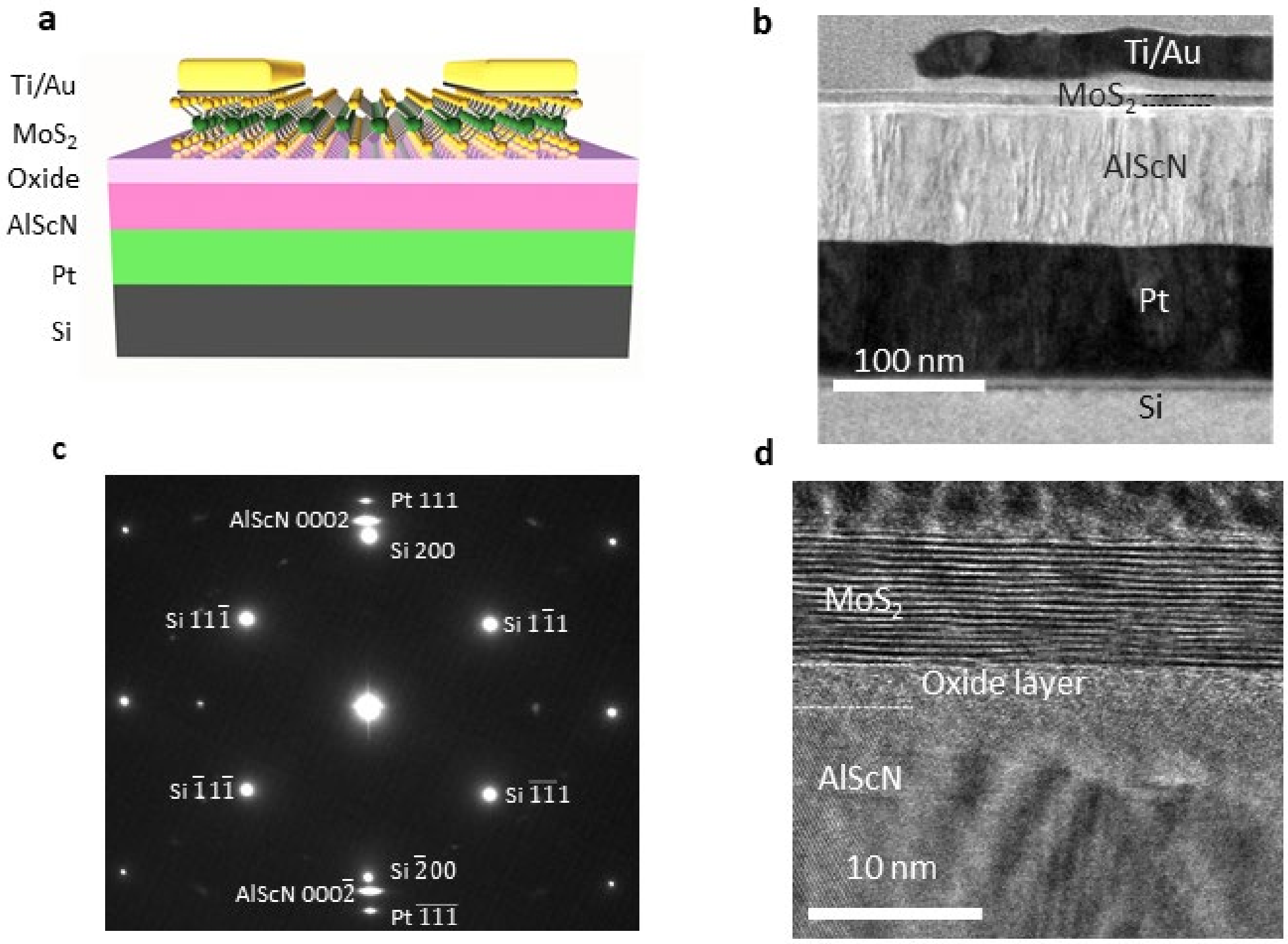

**Fig. 1. Multilayer $MoS_2$ FE-FETs with 100-nm-thick AlScN ferroeletric.** a, Schematic view of a AlScN/$MoS_2$ FE-FETs. The gate stack includes 100-nm-Pt on Si as the gate electrode, 100 nm AlScN as the ferroelectric gate dielectric. A ~3 nm native oxide layer naturally occurs on top of ferroelectric AlScN. b, Cross-sectional TEM image of a representative sample showing the source-

drain contact, $MoS_2$ channel, amorphous oxide and polycrystalline AlScN gate dielectric. c, SAED pattern of ferroelectric AlScN film showing templating along [0001](c-plane). d, High-resolution phase-contrast TEM image obtained from the channel/dielectric interface in c where the individual $MoS_2$ layers in the channel and surface oxide layer are visible.

**Performance of AlScN/$MoS_2$ FE-FETs**

The ferroelectric response of the 100 nm AlScN thin film was characterized by a positive-up, negative-down (PUND) measurement using a 10 µs square wave with a 1 ms delay between the two pulses (Supplementary Fig. S4a), as shown in Fig. 2a. The PUND test was preferred over a polarization-electric field hysteresis loop (P-E loop) because the P-E loop of 100 nm AlScN shows a polarization dependent leakage which hinders the observation of polarization saturation in the positive field side, as described in Fig. S4c and ref. [19]. The PUND result indicates a remanent polarization ~80 $\mu C/cm^2$ in the AlScN film and the onset of ferroelectric switching is over 50 V for a 10 µs pulse. Since the coercive voltage of a ferroelectric thin-film can be associated with the frequency (or pulse width) of the applied voltage [7, 20-22], another PUND test using monopolar triangular wave with a 100 ms pulse width and a zero-to-peak voltage of 40 V was conducted and is presented as the inset of Fig. 2a (See also Supplementary Fig. S4b). This PUND measurement indicates the same remanent polarization ~80 $\mu C/cm^2$ and a switching voltage between 30-40 V under the conditions similar to the device measurement and simulations. A non-volatile resistive switching at an applied DC voltage of 30-40 V was also observed, which serves as complementary evidence in support of ferrolectric switching (Supplementary Fig. S5).

In addition to standard current–voltage measurements, hysteresis between transfer characteristics was measured for the FE-FETs along two different sweep directions: i. forward (from low to high current i.e. negative to positive gate voltage) and ii. reverse (from high to low current and positive to negative gate voltage) gate voltage sweeps at two different drain voltages (Figure 2b). The application of a positive gate bias results in a sharp increase of drain current by several orders of magnitude associated with strong electron accumulation and a remanent on-state current after bringing the bias back to the read voltage window. The transistor shows an n-type characteristic due to the n-doping induced by the presence of sulfur vacancies in $MoS_2$ [24]. Upon switching from positive bias to negative gate bias, a gentle decline in current followed by a precipitous drop at ~-30 V is observed. This suggests channel depletion and a remanent off-state current. The drain current after reaching depletion (off-state) closely mimics the gate current, which indicates that it is due to leakage from the gate-insulator. The measured on/off current ratios of the reported devices are ~$10^6$.

The hysteresis in the reported transfer curves is significant and counterclockwise in its directionality, consistent with theory [25-28]. The significant counterclockwise hysteresis loops measured in AlScN/$MoS_2$ FE-FETs indicate that the ferroelectric polarization is dominating channel conduction. Charge trapping in dielectrics is also known to produce hysteresis loops in transfer curves. However, trapping-induced hysteresis loops are only clockwise in their directionality. Hence, our observation of a counterclockwise ferroelectric loop suggests the existence of ferroelectricity and polarization switching. With 100-nm-thick AlScN as a ferroelectric dielectric, the resulting AlScN/$MoS_2$ FE-FETs exhibit a large memory window ~ 35 V. The normalized memory window at a drain voltage of 1 V, divided by the thickness of the ferroelectric film, is 0.3 V/nm. To the best of our knowledge, this is the largest value reported for a FE-FET at room temperature. A large normalized memory window is critical for ultimate scaling and integration into very large scale circuits. A smaller memory window would result in increased bit flip errors and retention issues upon aggressive thickness scaling of the dielectric. This large normalized memory window is attributed to the desirable ferroelectric properties of AlScN: large coercive field, high polarization and moderate dielectric constant. We also performed technology computer-aided design (TCAD) simulations to investigate the dependence of memory window on the ferroelectricity of the gate dielelctric, by adoping the FE parameters of traditional Lead Zirconium Titanate (PZT), doped Hafnium Oxide ($HfO_2$), and AlScN (Supplementary Table 1). As shown in the simulation results in Fig. 2c, counterclockwise transfer curve hysteresis loops are observed in all three FE-FETs with $MoS_2$ channels and with 100 nm thick gate dielectrics. The simulated AlScN/$MoS_2$ FE-FET stands out for its large memory window when compared to alternative FE materials.

To further characterize the channel properties, output characterisitics (drain current-drain voltage) of AlScN/$MoS_2$ FE-FETs were measured. They show a linear behavior with a large degree of current control (current ratio of $10^4$) up to a high drain bias of 5 V and a large drive current density of over 100 μA/μm in the ON State. This ON current density is comparable with some of the highest current density values reported using electrical double layer dielectrics [29] or ultrathin atomic layer deposition grown dielectrics [30]. A maximum drain current over 1.4 mA is achieved with 100-nm-thick AlScN ferroelectric and a channel length of 3.2 μm. To further determine the memory effect and reliability, we performed cycling and time varying retention tests between the ON/OFF states as shown in Figure 2 e, f. Fig. 2e presents remanent ON-state and OFF-state currents extracted from 20 manually performed DC cycles. Cyclic program/erase operations of the same AlScN/$MoS_2$ FE-FET indicate that the both ON- and OFF- current states are stable and rewritable. Readouts at various delay times were carried out to determine retention (Fig. 2f). The low and high current/resistance states can be retained for at least 10,000 secs at room temperature without obvious degradation. The above reliability and retention characteristics hold promise for future NVM applications in a highly scaled and low-power device format. We emphasize that a 2D channel is critical in achieving not only a high ON/OFF ratio but also long retention due to minimal depolarization field [18].

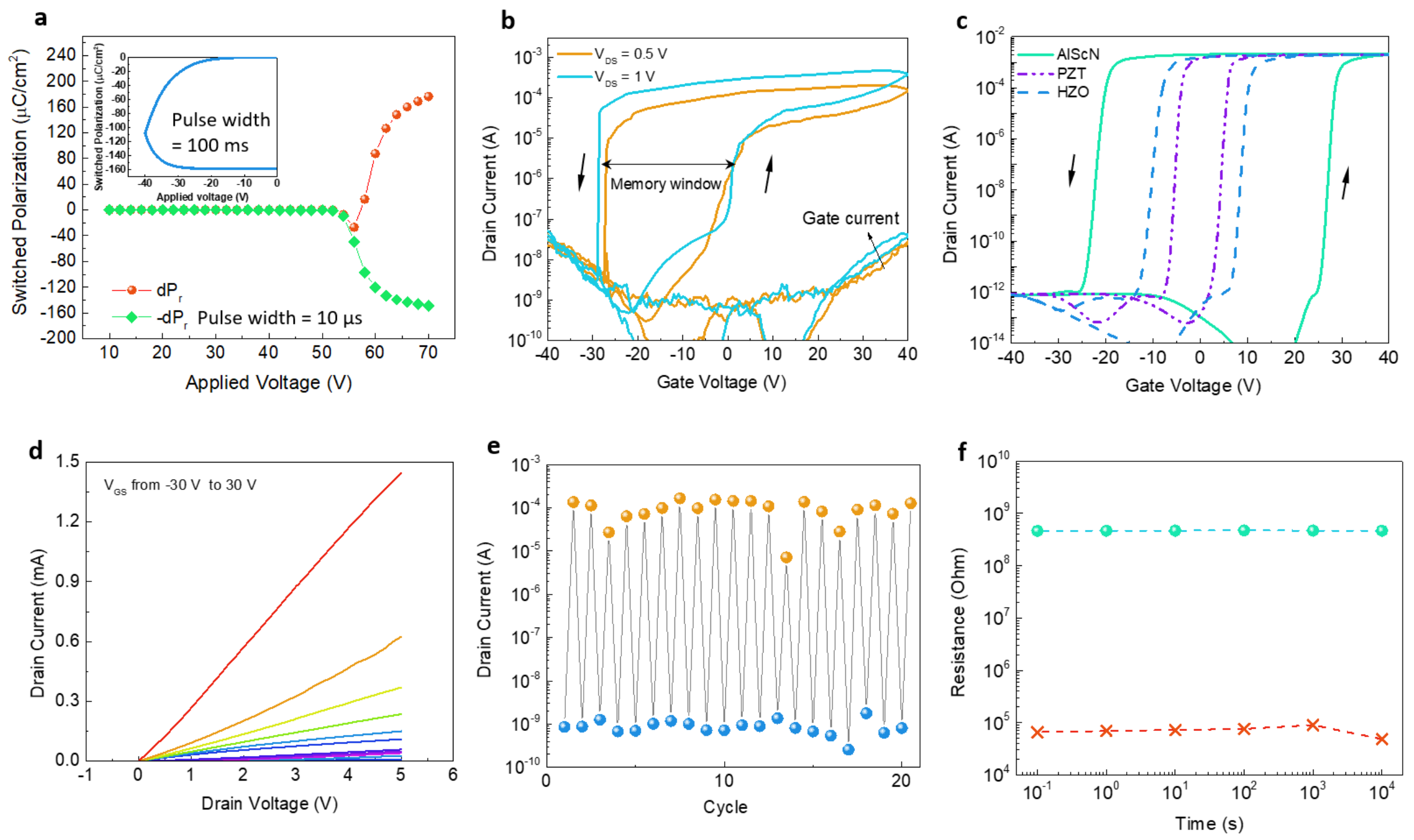

**Fig. 2. Room-temperature electrical characterization of AlScN/$MoS_2$ FE-FETs**. a, The PUND results of a 100 nm AlScN thin film with a pulse width of 10 µs and 1 ms delay between pulses. The PUND test reveals a saturated remanent polarization of 80 µC/$cm^2$ and a ferroelectric switching voltage over 50V. The inset is a monopolar triangular wave PUND with 100 ms pulse width and a zero-to-peak voltage of 40V, suggesting a ferroelectric switching voltage between 30 V - 40 V. b, Semi-logarithmic scale transfer characteristics at room temperature of a representative AlScN/$MoS_2$ FE-FET with 100 nm AlScN as ferroelectric gate dielectric. The channel length is 3.2 µm and the channel width is 14.8 um for the device shown. The arrows show the clockwise hysteresis of the drain current, which is consistent with accumulation and depletion of electrons. The device exhibits a large memory window (~30 V) (counterclockwise hysteresis loop) and on/off drain-to-source resistance ratio up to $10^6$. C, TCAD simulations of $MoS_2$ FE-FETs with ferroelectric dielectric of 100-nm-thick PZT, $HfO_2$, and AlScN, respectively showing counterclockwise transfer curve hysteresis loops of. d, Linear-scale output characteristics of the same device at various gate voltages. e, The ON- and OFF-state drain current as a function of the number of applied DC gate voltage sweeps. The gate voltage sweep ranges were from -40 V to +40 V. f, Resistance state retention measurement obtained by programming the ON- or OFF-state with a gate voltage of ±40 V, then monitoring the drain current for varying time intervals up to 10,000 secs. No significant degradation was observed on both ON- and OFF-state drain current.

### Comparison of AlScN/$MoS_2$ FE-FETs with reference AlN/$MoS_2$ FETs

To get a more direct evidence and further reinforce our observations from electrical measurements, we performed electrical measurement on reference AlN/$MoS_2$ FETs. An AlN film of identical thickness to AlScN was sputtered in the same pulsed DC Physical Vapor Deposition system under similar conditions. Then AlN/$MoS_2$ FETs with similar channel thickness and channel dimension were fabricated using the same process flow as the AlScN/$MoS_2$ FE-FETs. The dielectric constant (k) of AlN was measured to be ~7.1, which is comparable to the value of AlScN ~ 12. Transfer curves on both AlScN/$MoS_2$ FE-FETs and AlN/$MoS_2$ FETs with various sweep ranges are shown in Fig. 3. The hysteresis loops in the transfer characteristic, which depend strongly on gate voltage sweep range, serve as a strong evidence of ferroelectric polarization switching in AlScN/$MoS_2$ FE-FETs for DC measurements. As shown in Fig. 3a, the hysteresis in the AlScN/$MoS_2$ FE-FETs transfer curve is observed to reverse from clockwise to counterclockwise with the onset of polarization switching, which occurs at a higher gate voltage. For a small sweep range, the hysteresis is observed to be clockwise, indicating charge trapping is dominating the observed current hysteresis and that the ferroelectric polarization has not been reversed. For a large sweep range, the hysteresis is observed to be counterclockwise, indicating onset of polarization switching of the ferroelectric. As shown Fig. 3b, $MoS_2$ FETs on 100 nm AlN show regular n-type behavior with ON/OFF ~ $10^5$ similar to those made on oxide dielectrics. Significant hysteresis has also been observed, comparable to oxide dielectrics, but in the clockwise direction only. We attribute this to the trapped charge in the defects and dangling bonds in the dielectrics. Moreover, the hysteresis loop direction for the AlN dielectric FETs does not flip sign upon extension of the sweep range up to ±70 V (Supplementary Fig. S6). This further suggests that the counterclockwise hysteresis in AlScN/$MoS_2$ FE-FETs is induced by ferroelectric polarization switching.

Another striking observation is the difference in the magnitude of the ON current between the AlN/$MoS_2$ FETs and the AlScN/$MoS_2$ FE-FETs for similar channel dimensions, thicknesses and dielectric constants and dielectric thicknesses (Figure 3c). Given that the k of AlScN is only ~1.7 X greater than AlN, a difference in ON current density by 400 X is inexplicable by the standard dielectric capacitive charging model. This again suggests the presence of a high surface charge density in the semiconductor channel, which is induced by the ferroelectric polarization of AlScN. Beside the current magnitude, the comparative shift in threshold voltages of the transfer characteristics between AlN/$MoS_2$ FETs and AlScN/$MoS_2$ FE-FETs upon forward and reverse sweeps also suggests the presence of ferroelectric polarization in AlScN (Figure 3c). For the forward sweeps (red), the shift in threshold voltage between AlScN/$MoS_2$ FE-FETs and AlN/ $MoS_2$ FETs is 24 V, indicating additional negative charge at AlScN/$MoS_2$ interface. On the contrary, for the reverse sweeps (blue), the shift in threshold voltage between AlScN/$MoS_2$ FE-FETs and AlN/$MoS_2$ FETs is -19 V, indicating additional positive charge at the AlScN/$MoS_2$ interface. This change in sign of the charge at the AlScN/$MoS_2$ interface further verfies that the significant hysteresis induced in the transfer characteristics is a result of ferroelectric polarization switching. Finally, the dissimilarity of the subthreshold swing slope (SS) between AlN/ $MoS_2$ FETs and AlScN/$MoS_2$ FE-FETs also indicates that the density of carriers in the channel is dominated by ferroelectric polarization switching. As shown in Fig.3(c), the demonstrated $MoS_2$ FETs on AlScN exhibits steep slope-switching behavior (100 mV/dec) compared to the $MoS_2$ FETs on AlN of the same thickness (3000 mV/dec), reducing the subthreshold swing by 2 orders of magnitude. The abrupt and dramatic switching behavior displayed in the Fig.3(c) is a direct effect of negative capacitance by integrating a ferroelectric layer into the gate [31-33]. In the negative capacitance effect, the insulating ferroelectric layer serves as a negative capacitor during the ferroelectric switching process such that the channel surface potential gets amplified more than the gate voltage. A consequence of this phenomena is that the FET device can operate with SS less than 60 mV/dec at room temperature in addition to exhibiting negative drain-induced barrier lowering. We have observed both of these effects in our devices (Supplementary Figs. S7-8) which further verifies our claim of FE switching.

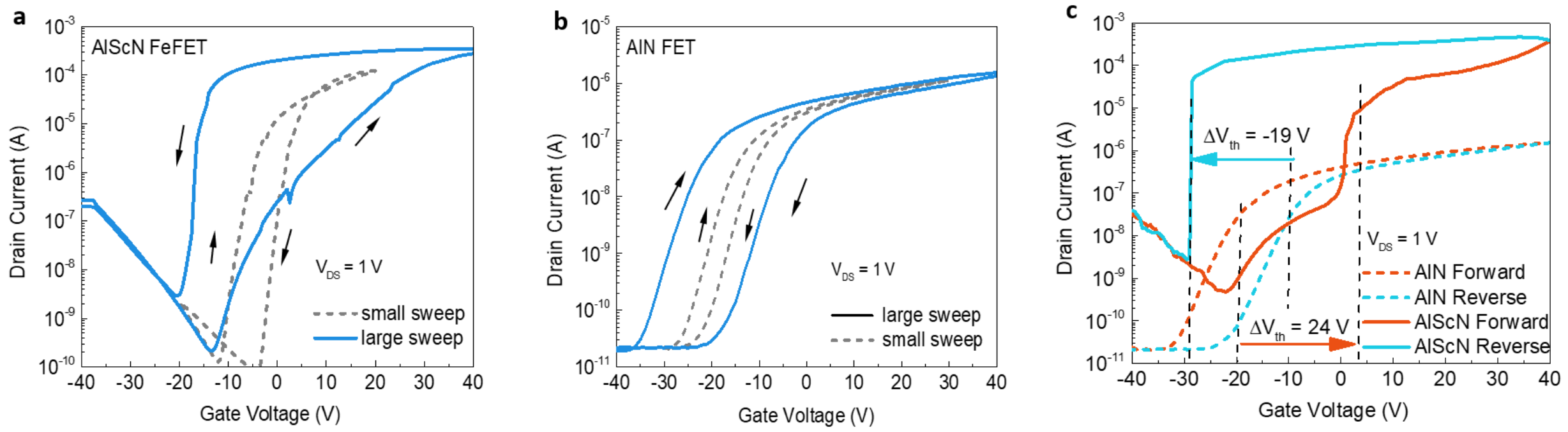

**Fig. 3. Comparison of AlScN/$MoS_2$ FE-FETs with reference AlN/$MoS_2$ FETs.** a, Semilogarithmic scale transfer characteristics at room temperature of a representative AlScN/$MoS_2$ FE-FETs with a large gate voltage sweep range (solid blue line) and a small gate voltage sweep range (dashed grey line). The device has an AlScN thickness of 100 nm, a channel length of 3.75 µm and channel width of 15 µm. The arrows show the hysteresis directionality of the transfer curves. For a small sweep range (from -20 V to 20 V), the hysteresis is observed to be clockwise, indicating charge trapping and no ferroelectric switching. Conversely, for a large sweep range, the hysteresis is observed to be counterclockwise, indicating the occurance of FE polarization switching. b, Semi-logarithmic scale transfer characteristics at room temperature of a representative AlN/$MoS_2$ FET with both a large gate sweep range (solid blue line) and a small gate sweep range (dash grey line). The device has an AlN gate thickness of 100 nm, a channel length of 3.6 µm and channel width of 10 µm. For both small and large gate voltage sweep ranges, the hysteresis is observed to be clockwise, indicating that charge traps dominate the threshold voltage shifts and no ferroelectricity is present. c. Comparative transfer characteristics during both forward and back ward sweeps in the +- 40 V range for AlScN and AlN dielectric $MoS_2$ FETs. The clear and opposite signs of threshold voltage shifts between AlScN/$MoS_2$ FE-FET and AlN/$MoS_2$ FET for forward (red) and reverse (blue) sweeps is shown. The large differences in ON currents between the two

transistors and the marked enhancement in steepness of the sub-threshold swing in AlScN based $MoS_2$ FETs as compared AlN based FETs is also evident.

**Benchmarking and Discussion**

The above results show that FE AlScN and 2D $MoS_2$ channel-based FE-FETs have appealing electrical characteritics and memory device performance. However, a thorough benchmarking with available FE-FETs in the literature is desired to enable a fair assessment and viability for a future technology. Figure 4a shows a comparison of AlScN/$MoS_2$ FE-FETs with previously reported FE-FETs including traditional PZT FE-FETs and the state-of-the-art $HfO_2$ based FE-FETs. Si channels, 2D $MoS_2$ channels, and other ultrathin body channel materials have been used for the comparison. We have extracted the on-state current to off-state current ratio and a normalized memory window (memory window/ferroelectric thickness) from the reported transfer curves for a fair benchmarking. Normalization of the memory window to FE thickness is essential for the comparison, since all FE materials used and reported in the litereature for FE-FET memory applications have varying thicknesses and because the memory window is propotional to the product of ferroelectric thickness and coercive field. The comparison data clearly indicate that AlScN/$MoS_2$ FE-FETs outperform all previous FE-FETs in terms of normailized memory window and, more importantly, AlScN/$MoS_2$ FE-FETs fullfill the two critical requirements (sufficient memory window and high ON/OFF current ratio) concurrently. Although the larger coercive field implies higher write fields in theory, the shape of P-E loops of the FE dielectrics must be taken into account together with the coercive fields to determine the write fields. For ferroelectrics with non-ideal P-E loops such as PZT and $HfO_2$, a write field much higher than the coercive field is usually needed to saturate the polarization states. The more square shaped (near-ideal) P-E loops of AlScN are expected to relieve the requirements of higher write fields as compared to the coercive fields. Besides the non-ideal P-E behavior, traditional perovskite oxide ferroelectrics such as lead zirconium titanate (PZT) or strontium bismuth tantalate (SBT) are incompatible with the BEOL, CMOS processing temperatures (< 350 °C). Further, lead and bismuth are volatile components that are likely to diffuse and contaminate the whole chip upon experiencing such temperatures during operation or fabrication. Among oxide FE materials discovered to date, only doped $HfO_2$ has been a serious contender, with the potential to be adoped in CMOS BEOL applications like monolithic 3D integration of memory interleaved with logic computation layers. However, a high temperature annealling (> 400 °C) is necessary for most doped $HfO_2$ to crystalize and attain decent ferroelectricity.[ref needed] In comparison, AlScN can be deposited in a polycrystalline FE switching ready state at 350 °C on top of Si wafers as we have shown. Further, near room-temeprature transfer of wafer scale CVD grown 2D semiconductors has also been achieved.[ref] Therefore the device concept presented has the highest potential to be ultilzed in CMOS BEOL applications among all the FE-FETs.

While AlScN with a 2D chalcogenide channel Fe-FET is certainly appealing in terms of ON/OFF ratio and memory window, the endurance and retention are equally critical for a memory application. One constraint/draw back that the FE-FET device concept has traditionally encountered is the depolarization of the ferroelectric layer over time due to the electrical field induced by incomplete charge compensation of the semiconducting channel. By assuming the same amount of incompletely compensated charge, the depolarization field in the ferroelectric is approximately inversely proportion to its thickness. Thus, theoretically, as the ferroelectric layer thickness scales down for low-voltage operation, the remanent polarization states will become unstable as the depolarization field becomes larger than the coercive field. This depolarization field has been the cause of severe retention loss in PZT and SBT based FE-FETs, which has hindered further development of FE-FET memory on perovskite oxide ferroelectrics in practical applications [18, 44]. To investigate depolarization field in the proposed AlScN FE-FET and its counterparts (PZT and $HfO_2$ FE-FET), we calculate the depolarization field/coercive field ($E_{dep}/E_c$) ratio in the FE-FETs using a short circuit model together with the ferroelectric and dielectric parameters ($E_c$, $P_r$ and k) that describe the ferroelectrics. A 1-nm-thick interfacial insulator layer is assumed to model realistic depolarization properties in a FE-FET. For fair comparisons, the same remnant polarization value of 30 $\mu C/cm^2$ has been used in the calculation of the $E_{dep}/E_c$ ratio. This is because the memory window of the AlScN FE-FET saturates beyond this polarization (Supplementary Fig. S8). As shown in Figure 4b, even for a scaled 10-nm AlScN ferroelectric layer, we derive an $E_{dep}/E_c$ ratio less than 1, indicating that the depolarization field does not exceed the coercive field for the ferroelectric layer. In contrast both $HfO_2$ and PZT based FE-FETs have $E_{dep}/E_c$ ratio of ~3 and ~40 respectively. This suggests that low voltage scaling of AlScN/2D channel-based FE-FETs concurrent with long retention is attainable. Furthermore, ferroelectricity in < 10 nm AlScN has already been demonstrated and the lateral scaling advantages for 2D channels are already well-proven [45-46].

In summary, we have shown high performance Fe-FET based memory devices using AlScN FE dielectric combined with a 2D $MoS_2$ channel that exhibit a record normalized memory window and ON/OFF ratio concurrently with good retention and CMOS BEOL compatible processing temperatures. Thus, our results pave a path towards a practical BEOL CMOS compatible memory technology.

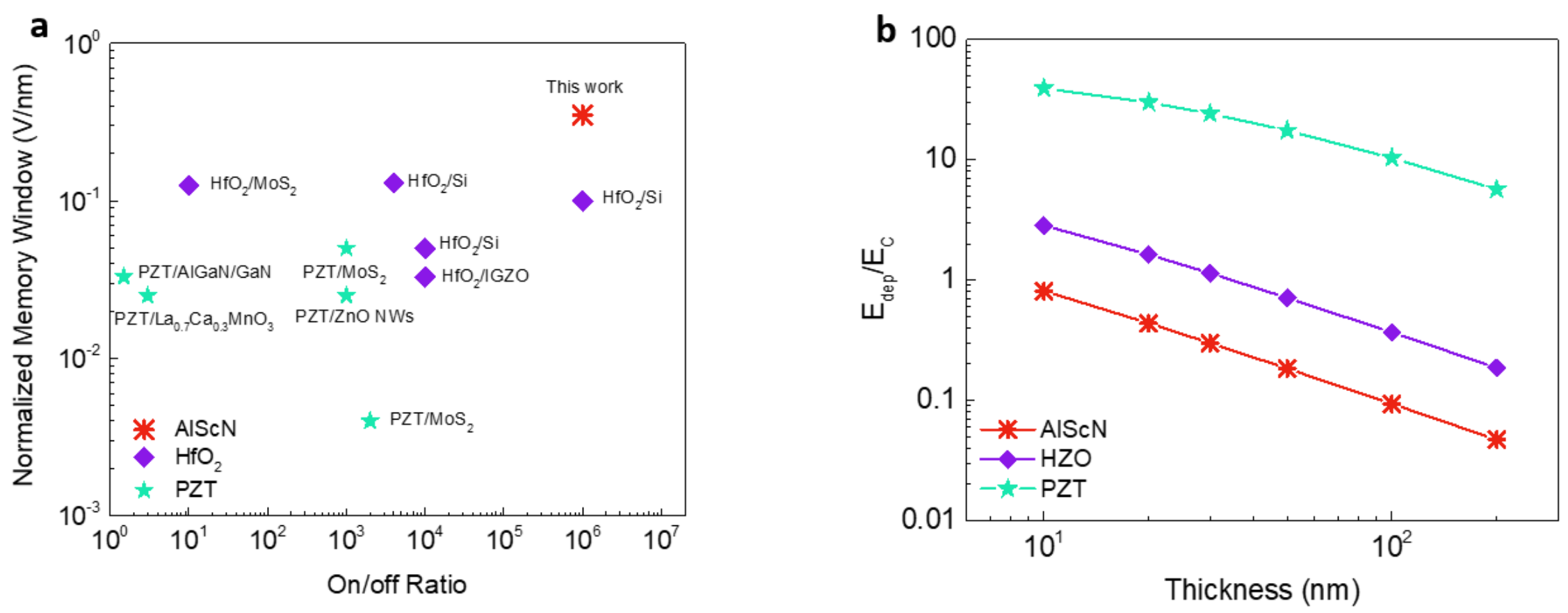

**Fig. 4. Performance comparison of the AlScN/$MoS_2$ FE-FETs with previously reported FE-FETs including tradional PZT FE-FETs and the state-of-the-art $HfO_2$ FE-FETs.** a, ON/OFF current ratios and normalized memory window from the reported Fe-FETs in the literature with HZO and PZT dielectrics paired with Si and 2D semiconductor channels. It is worth noting that the reported AlScN -based FE-FET lies closest to the top-right corner of the plot. b, The calculated ratio of depolarization field over coercive field ($E_{dep}/E_c$) in three different FE-FET cases: 1) AlScN + 1 nm buffer layer, 2) $HfO_2$ + 1 nm buffer layer, and 3) PZT + 1 nm interfacial insulating layer shown as a function of varying thickness of the FE dielectric. $E_{dep}/E_c$ in AlScN FE-FETs is much smaller than its $HfO_2$ and PZT FE-FET counterparts with $E_{dep}/E_c$ ratio less than 1 even when scaled to 10 nm thicknesses.

## METHODS

### Device fabrication

We start by sputtering ferroelectric AlScN/gate stack on the top of Si wafer. A 100-nm-thick Pt film was deposited by sputtering onto a Ti seed on the top of Si (100) substrate. Next, a 100-nm thick ferroelectric AlScN film was co-sputtered from two separate 4-inch Al (1000W) and Sc (450W) targets in an Evatec CLUSTERLINE® 200 II pulsed DC Physical Vapor Deposition System [19]. The deposition was done at 350 °C with 6 sccm of Ar and 12 sccm of $N_2$ gas flow. This yields a stress of -569 MPa and a surface roughness of 1.3 nm (Supplementary Fig. S2). For conventional AlN, the deposition was done at 350 °C with only 20 sccms of $N_2$ gas flow without utlization of the Sc target. A few-layer $MoS_2$ flake was mechanically exfoliated from bulk $MoS_2$ crystals (HQ graphene) using Scotch tape and then transferred onto a PDMS stamp. An optical microscope was used to locate uniform and thin $MoS_2$ flakes on the PDMS stamp. $MoS_2$ flakes were then dry-transferred onto ferroelectric AlScN or AlN substrates using the PDMS stamp as transfer medium. Electrical contacts were patterned onto $MoS_2$ flakes using standard electron-beam lithography. Firstly, PMMA e-beam resist (MicroChem 495 A8) was spin-coated (4000 rpm for 80 seconds) onto the the p++ Si/Pt/AlN or p++Si/Pt/AlScN wafers followed by baking on a hot plate at 180 °C for 10 min. Second, we used e-beam lithography to define the source/drain regions in the PMMA layer followed by development using MicroChem's developer (MIBK:IPA (1:3)). The patterned samples were then cleaned using an $O_2$ plasma (10 s at 50 W) before the metal electrode deposition step to remove any potential residues, followed by e-beam evaporation of Ti/Au film (10 nm/40 nm) and lift-off processes.

### Device characterization

Electrical measurements were performed in air at ambient temeprature in a Lakeshore probe station using a Keithley 4200A semiconductor characterization system. AFM characterizaitons were performed using an AIST-NT SPM SmartSPM™-1000 which was done in the tapping mode with a 200 kHz resonance frequency. P-E hysteresis loops and PUND measurements of ferroelectric AlScN were conducted using a Radiant Precision Premier II (Radiant technologies, Inc., Albuquerque, NM) testing platform. The cross-section TEM sample was prepared in a FEI Helios Nanolab 600 focused ion beam (FIB) system using the in-situ lift-out technique. The sample was first coated with a thin carbonaceous protection layer by writing a line on the surface with a Sharpie® marker. Subsequent electron beam and ion beam deposition of Pt protection layers were used to prevent charging and heating effects during FIB milling. At the final cleaning stage, a low-energy Ga+ ion beam (5 keV) was used to reduce FIB-induced damage. TEM characterization and image acquisition were carried out on a JEOL F200 TEM operated at 200 kV accelerating voltage. Image analysis and feature extraction were performed using ImageJ. All quantification results presented in this work were calculated with Digital Micrograph software (DM, Gatan Inc., USA). Physics-based self-consistent FE-FETs current-voltage characteristics simulation was performed using Synopsys Sentaurus TCAD, by coupling Poisson's equation, the Ginzburg-Landau equation and the 2D charge conservation equation.

**Acknowledgements**

This work was primarily supported by the DARPA TUFEN program. The work was carried out in part at the Singh Center for Nanotechnology at the University of Pennsylvania which is supported by the National Science Foundation (NSF) National Nanotechnology Coordinated Infrastructure Program (NSF grant NNCI-1542153). J.M. was partially supported by a grant from the Air Force Office of Scientific Research (20RT0130). The authors gratefully acknowledge use of facilities and instrumentation supported by NSF through the University of Pennsylvania Materials Research Science and Engineering Center (MRSEC) (DMR-1720530). TEM sample preparation was performed by Kim Kisslinger at the Center for Functional Nanomaterials, Brookhaven National Laboratory, which is a U.S. DOE Office of Science Facility, at Brookhaven National Laboratory under Contract No. DE-SC0012704.

**Author Contributions**

D.J., R.O., E.S. and X.L. conceived the idea. X.L. designed and made the memory devices, also performed electrical measurements. X.L. performed all simulations and modelling. D.W performed P-E loop and PUND measurements. J.Z performed ferroelectric material growth. P.M. performed transmission electron microscopy characterization. J.M. assisted with sample preparation and microfabrication. D.J., R.O., E.S. and X.L. analyzed and interpreted the data. D.J., R.O and E.S. supervised the study. All others contributed to writing of the manuscript.

**Competing interests**

The authors declare the following competing interests: D.J., X.L., R.O. and E.A.S. have a provisional patent filed based on this work. The authors declare no other competing interests.

**Additional information**

Supplementary information is available for this paper at:
Reprints and permissions information is available at www.nature.com/reprints.
Correspondence and requests for materials should be addressed to

Supplementary materials for this article is available at

Table. S1. Benchmarking and comparison of the ferroelectric AlScN with tradional perovskite oxide ferroelectric PZT and ferroelectric $HfO_2$.
Fig. S1. Optical micrographs of AlScN/$MoS_2$ FE-FET devices.
Fig. S2. AFM topography image of a low temperature deposited AlScN ferroelectric across 4-inch wafer.
Fig. S3. TEM characterization of a AlScN/$MoS_2$ FE-FETs.
Fig. S4. Ferroelectric response in 100 nm AlScN.
Fig. S5. Non-volatile resistive switching of an AlScN ferroelectric MIM device diode.
Fig. S6. Room-temperature electrical characterization of a AlN/$MoS_2$ FET.
Fig. S7. Steep sub-threshold slope induced by negative capacitance through integrating ferroelectric AlScN in the gate stack.
Fig. S8. Negative drain-induced-barrier-lowering (DIBL) induced by negative capacitance in the gate stack.
Fig. S9. TCAD simulation of AlScN/$MoS_2$ FeFTEs with various remanent polarizations.

**References**

[1] J. L. Moll and Y. Tarui. A new solid state memory resistor. *IEEE Trans. on Electron Devices*, **10**, 338-338, (1963).

[2] J. S. Meena, S. M. Sze, U. Chand, and T.-Y. Tseng. Overview of emerging nonvolatile memory technologies. *Nanoscale Research Lett.*, **9**, 526 (2014).

[3] Y. Arimoto and H. Ishiwara. Current status of ferroelectric random-access memory. *MRS Bull.*, **29**, 823-828, (2004).

[4] D. C. Yoo et al. Highly reliable 50nm-thick PZT capacitor and low voltage FRAM device using Ir/SrRuO3/MOCVD PZT capacitor technology. In *Symposium on VLSI Technology* 100-101 (IEEE, 2005).

[5] Ishiwara, H. FeFET and ferroelectric random access memories. J. *Nanosci. Nanotechnol*. **12**, 7619–7627 (2012).

[6] N. Setter et al. Ferroelectric thin films: Review of materials, properties, and applications. *J. Appl. Phys.* **100**, 051606, (2006).

[7] S. Fichtner, N. Wolff, F. Lofink, L. Kienle, and B. Wagner. AlScN: A III-V semiconductor based ferroelectric. *J. Appl. Phys*. **125**, 114103 (2019).

[8] Eeckhout, Lieven. Is moore's law slowing down? what's next?. *IEEE Micro* **37**, 4-5 (2017).

[9] Efnusheva, Danijela, Ana Cholakoska, and Aristotel Tentov. A survey of different approaches for overcoming the processor-memory bottleneck. *International Journal of Computer Science and Information Technology* **9**, 151-163 (2017).

[10] Xu, X. et al. Scaling for edge inference of deep neural networks. *Nat. Electron*. **1**, 216–222 (2018).

[11] Wong, H.-S. P. & Salahuddin, S. Memory leads the way to better computing. *Nat. Nanotechnol*. **10**, 191 (2015).

[12] Krizhevsky, A., Sutskever, I. & Hinton, G. E. ImageNet classifcation with deep convolutional neural networks. *Comm. ACM* **60**, 84–90 (2012).

[13] Sayal, A., Fathima, S., Nibhanupudi, S. S. T. & Kulkarni, J. P. All-digital time-domain CNN engine using bidirectional memory delay lines for energy-efcient edge computing. In *IEEE International Solid-State Circuits Conference* (ISSCC) 228–229 (IEEE, 2019).

[14] S. Yu and P. Chen. Emerging memory technologies: Recent trends and prospects. *IEEE Solid-State Circuits Magazine*, **8**, 43-56 (2016).

[15] Müller, J. et al. Nanosecond polarization switching and long retention in a novel MFIS-FET based on ferroelectric $HfO_2$. *IEEE Electron Device Lett.* **33**, 185–187 (2012).

[16] T. Mikolajick, U. Schroeder and S. Slesazeck. The Past, the Present, and the Future of Ferroelectric Memories. *IEEE Trans. on Electron Devices* **67**, 1434-1443 (2020).

[17] Florent, K. et al. Vertical ferroelectric $HfO_2$ FET based on 3-D NAND architecture: towards dense low-power memory. In *Proceedings of IEEE International Electron Devices Meeting* 2.5.1–2.5.4 (IEEE, 2018).
[18] T. P. Ma and Jin-Ping Han. Why is nonvolatile ferroelectric memory field-effect transistor still elusive? *IEEE Electron Device Lett*. **23**, 386-388 (2002).
[19] D. Wang. et al. "Ferroelectric C-axis Textured Aluminum Scandium Nitride Thin Films of 100 nm Thickness," *In Conf. of the IEEE Int. Freq. Cntrl. Symp. (IFCS) and the IEEE Int. Symp. on Applications of Ferroelectrics (ISAF)* 1-4, (IEEE, 2020).
[20] S. Fichtner, D. Kaden, F. TLofink and B. Wagner. A Generic CMOS Compatible Piezoelectric Multilayer Actuator Approach Based on Permanent Ferroelectric Polarization Inversion in $Al_{1-X}Sc_xN$. In *20th International Conference on Solid-State Sensors, Actuators and Microsystems & Eurosensors XXXIII 289-292*, (IEEE, 2019).
[21] Luo, Q., Cheng, Y., Yang, J. *et al.* A highly CMOS compatible hafnia-based ferroelectric diode. *Nat Commun* **11,** 1391 (2020).
[22] Wei Li. et al. Calculation of frequency-dependent coercive field based on the investigation of intrinsic switching kinetics of strained $Pb(Zr_{0.2}Ti_{0.8})O_3$ thin films. *J. Phys. D: Appl. Phys*. **44**, 105404, (2011).
[23] J. F. Scott. Models for the frequency dependence of coercive field and the size dependence of remanent polarization in ferroelectric thin films. *Integrated Ferroelectrics* **12**, 71-81, (1996).
[24] Radisavljevic, B., Radenovic, A., Brivio, J. *et al.* Single-layer $MoS_2$ transistors. *Nature Nanotech* **6,** 147–150 (2011).
[25] Si, M., Saha, A.K., Gao, S. *et al.* A ferroelectric semiconductor field-effect transistor. *Nat Electron* **2,** 580–586 (2019).
[26] Miller, S. L. & McWhorter, P. J. Physics of the ferroelectric nonvolatile memory field effect transistor. *J. Appl. Phys.* **72**, 5999–6010 (1992).
[27] Yurchuk, E. et al. Charge-trapping phenomena in $HfO_2$-based FeFET-type nonvolatile memories. *IEEE Trans. Electron Devices* **63**, 3501–3507 (2016).
[28] Naber, R., Tanase, C., Blom, P. *et al.* High-performance solution-processed polymer ferroelectric field-effect transistors. *Nature Mater* **4,** 243–248 (2005).
[29] S. Fathipour. et al. Record high current density and low contact resistance in $MoS_2$ FETs by ion doping. *In 2016 International Symposium on VLSI Technology, Systems and Application (VLSI-TSA)* 1-2, (IEEE 2016).
[30] Dominik Lembke and Andras Kis. Breakdown of High-Performance Monolayer $MoS_2$ Transistors. *ACS Nano* **6**, 10070–10075, (2012).
[31] Muhammad A. Alam, Mengwei Si, and Peide D. Ye. A critical review of recent progress on negative capacitance field-effect transistors. *Appl. Phys. Lett*. **114**, 090401 (2019).
[32] Si, M., Su, C., Jiang, C. *et al.* Steep-slope hysteresis-free negative capacitance $MoS_2$ transistors. *Nature Nanotech* **13,** 24–28 (2018).

[33] Felicia A. McGuire, Yuh-Chen Lin, Katherine Price, et al. Sustained Sub-60 mV/decade Switching via the Negative Capacitance Effect in $MoS_2$ Transistors. *Nano Lett*. **17**, 4801–4806, (2017).

[34] M. Trentzsch et al. A 28nm HKMG super low power embedded NVM technology based on ferroelectric FETs. In *Proceedings of IEEE International Electron Devices Meeting (IEDM)* 11.5.1-11.5.4, (IEEE, 2016).

[35] W. Xiao et al. Performance Improvement of $Hf_{0.5}Zr_{0.5}O_2$-Based Ferroelectric-Field-Effect Transistors With $ZrO_2$ Seed Layers. *IEEE Electron Device Lett*. **40**, 714-717, (2019).

[36] M. Jerry et al. Ferroelectric FET analog synapse for acceleration of deep neural network training. In *Proceedings of IEEE International Electron Devices Meeting (IEDM)* 6.2.1-6.2.4, (IEEE, 2017).

[37] Mo et al. for High-Density and Low-Power Memory Application. In *Symposium on VLSI Technology* T42-T43, (IEEE, 2019).

[38] Wui Chung Yap et al. Ferroelectric transistors with monolayer molybdenum disulfide and ultra-thin aluminum-doped hafnium oxide. *Appl. Phys. Lett*. **111**, 013103 (2017).

[39] Alexey Lipatov et al. Optoelectrical Molybdenum Disulfide ($MoS_2$)—Ferroelectric Memories. *ACS Nano* **9**, 8089–8098 (2015).

[40] Changhyun Ko et al. Ferroelectrically Gated Atomically Thin Transition - Metal Dichalcogenides as Nonvolatile Memory. *Adv. Mater.* **28**, 2923-2930 (2016).

[41] S. Mathews et al. Ferroelectric Field Effect Transistor Based on Epitaxial Perovskite Heterostructures. *Science* **276**, 238-240 (1997).

[42] L. Chen et al. Polarization Engineering in PZT/AlGaN/GaN High-Electron-Mobility Transistors. *IEEE Trans. on Electron Devices* **65**, 3149-3155, (2018).

[43] L. Liao et al. Ferroelectric Transistors with Nanowire Channel: Toward Nonvolatile Memory Applications. *ACS Nano* **3**, 700–706 (2015).

[44] X. Pan and T. P. Ma. Retention mechanism study of the ferroelectric field effect transistor. *Appl. Phys. Lett*. **99**, 01305 (2011).

[45] Shinnosuke Yasuoka, et al. Effects of deposition conditions on the ferroelectric properties of $(Al_{1-x}Sc_x)N$ thin films. *J. Appl. Phys*. **128**, 114103 (2020).

[46] Sujay B. Desai, et al. $MoS_2$ transistors with 1-nanometer gate lengths. Science **354**, 99-102 (2016).

Supplementary Materials for

# Post-CMOS Compatible Aluminum Scandium Nitride/2D Channel Ferroelectric Field-Effect-Transistor

Xiwen Liu,[a] Dixiong Wang,[a] Jeffrey Zheng,[a] Pariasadat Musavigharavi,[a,b] Jinshui Miao,[a] Eric A. Stach,[b] Roy H. Olsson III,[a] Deep Jariwala[a]*

[a] Electrical and Systems Engineering, University of Pennsylvania, Philadelphia, PA, USA

[b] Material Science and Engineering, University of Pennsylvania, Philadelphia, PA, USA

*Corresponding author: dmj@seas.upenn.edu

| Ferroelectrics | PZT | $HfO_2$ | AlScN |
|---|---|---|---|
| Remnant Polarization (μC/cm$^2$) | 20-30 | 1-30 | 80-115 |
| Coercive Field (MV/cm) | 0.05 | 1-2 | 2-5 |
| Dielectric Constant | 450-1250 | 15-25 | 10-20 |
| Max. Processing Temperature (℃) | > 600 | 450-1100 | < 400 |

**Table. S1. Benchmarking and comparison of the ferroelectric AlScN with tradional perovskite oxide ferroelectric lead zirconium titanate (PZT) and ferroelectric $HfO_2$.** AlScN shows large coercive fields, $E_c$, of 2-4.5 MV/cm, which enables scaling to thinner ferroelectric layers and smaller FeFET gate dimensions while maintaining a large memory window. When combined with the high remnant polarizations, this leads to scaling of the bit density and ensures a high retention time of FeFETs against depolarization field. The low deposition temperature below 350 ℃ of AlScN without the need for a high-temperature annealing allows for FeFET integration directly in a CMOS BEOL process.

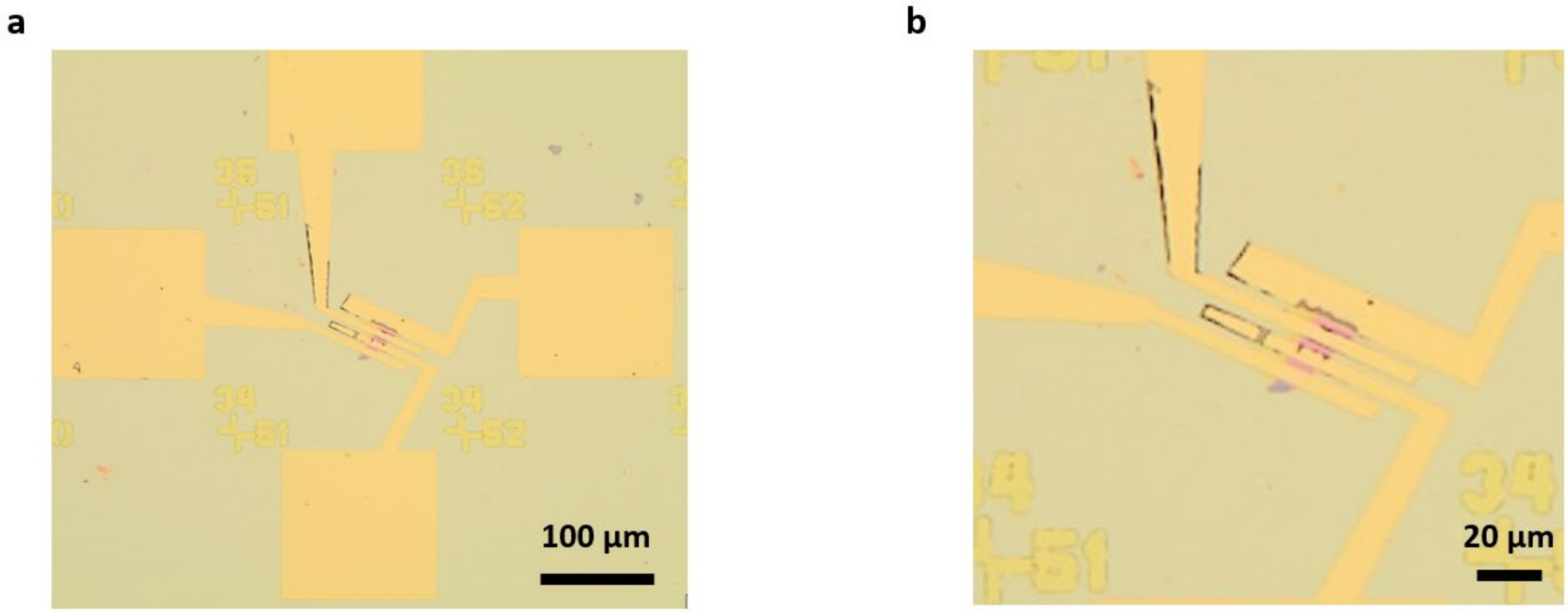

**Fig. S1. Optical micrographs of AlScN/$MoS_2$ FE-FET devices.**

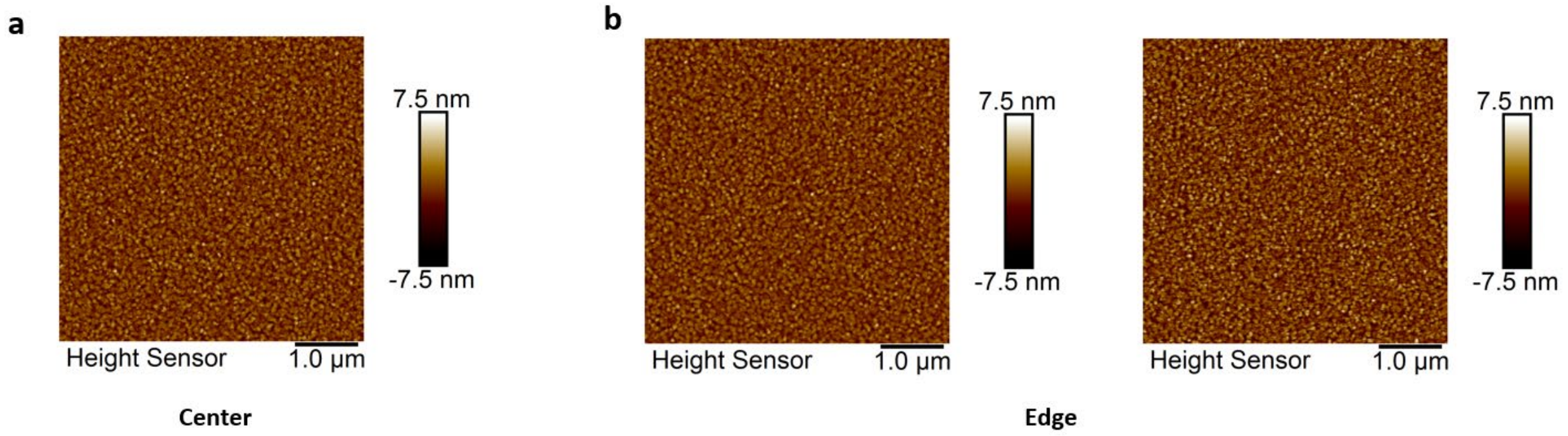

**Fig. S2. AFM topography image of a low temperature deposited AlScN ferroelectric across 4-inch wafer**. a-b, Center and edge AFM images across the 4-inch wafer, showing uniform surface morphology and small surface roughness of 1.3 nm over the full wafer.

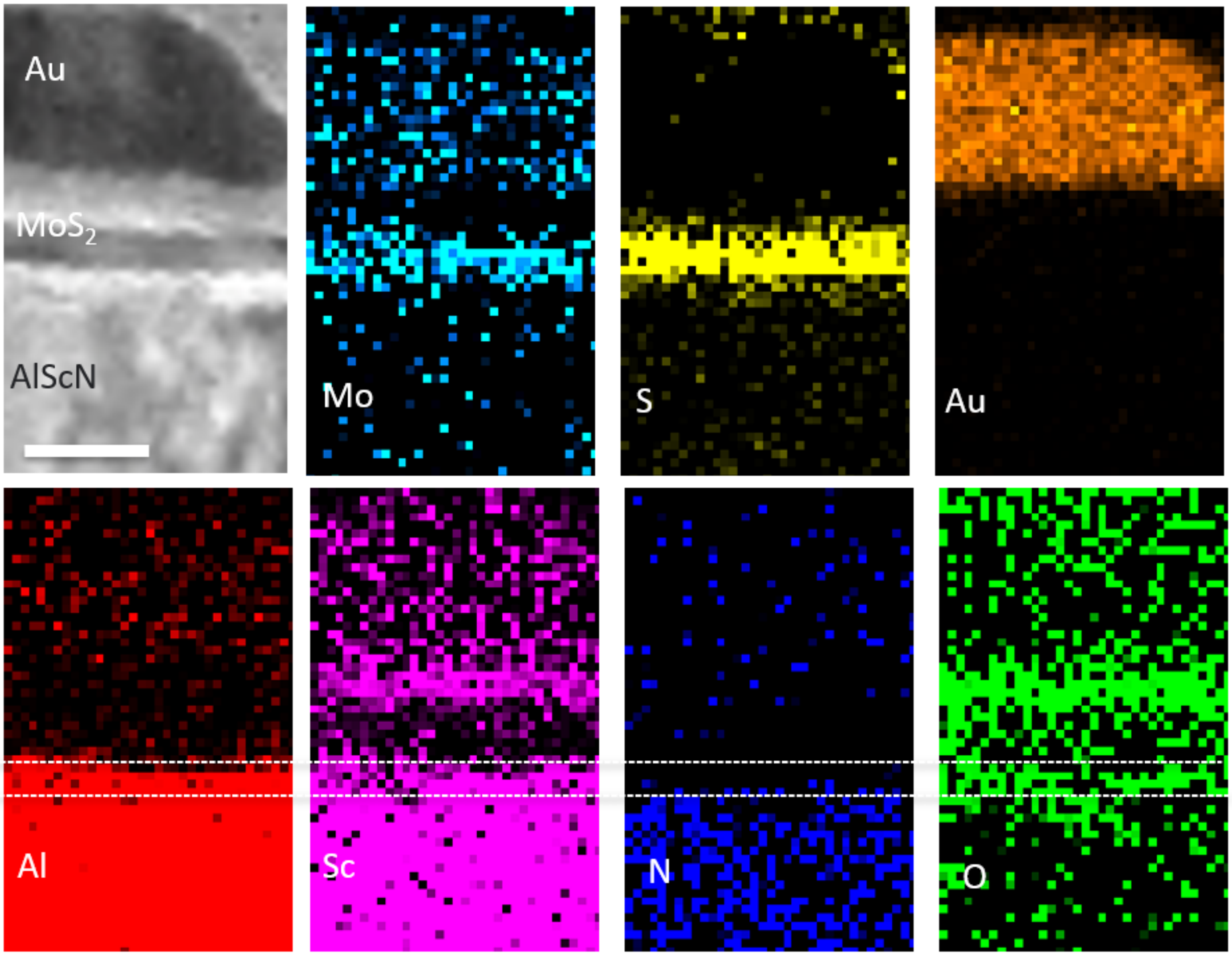

**Fig. S3. TEM characterization of a AlScN/$MoS_2$ FeFETs**. a, Cross-sectional TEM image of a representative AlScN/$MoS_2$ FeFETs showing AlScN gate ferroelectric, $MoS_2$ and Ti/Au source/drain electrode. Scale bar, 20 nm. b, Electron dispersive spectroscopy (EDS) map of the transistor region showing spatial distribution of Sc, N, Al, O, Mo and S elements, thus confirming the location of the AlScN and $MoS_2$ in the device. The oxide layer is indicated in between dashed lines, demonstrating Nitrogen atoms are replaced by Oxygen atoms.

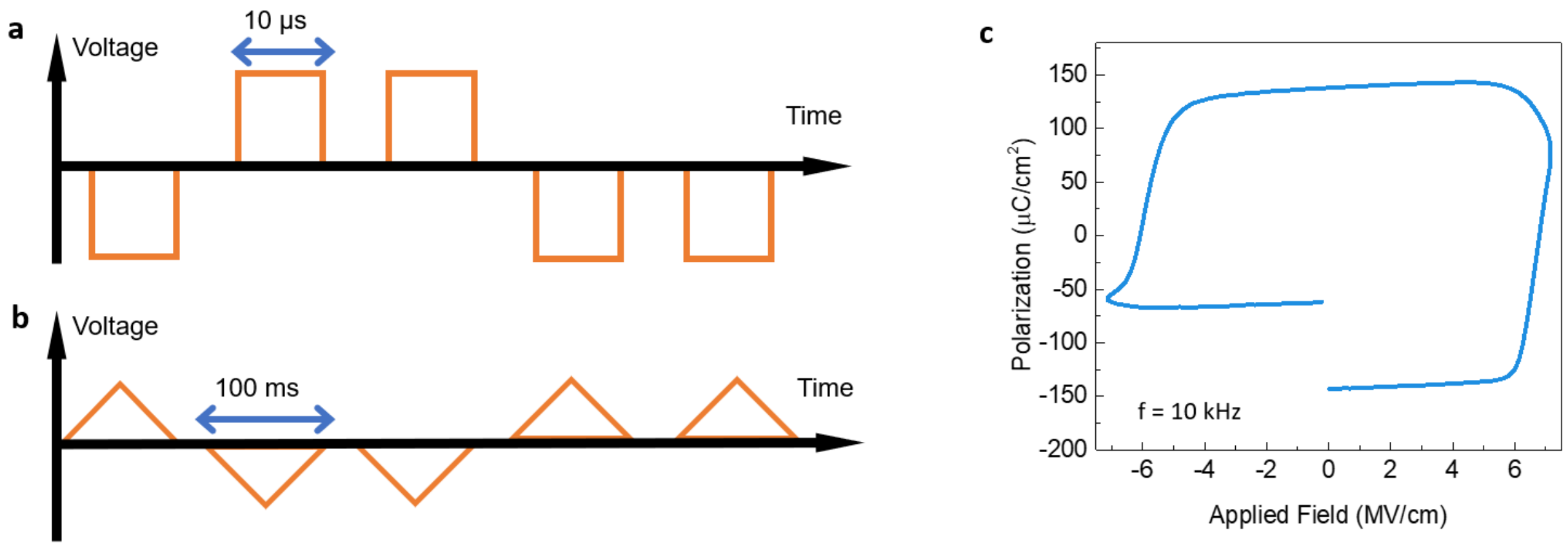

**Fig. S4. Ferroelectric response in 100 nm AlScN**. a-b, Schematics of the signal sequences for the PUND measurements to differentiate the ferroelectric and non-ferroelectric contributions to the polarization, corresponding to Fig. 2a and inset. c, The P-E hysteresis loops of a 100 nm thick AlScN on [111] at 100 kHz. The hysteresis loops are saturated in the negative field half of the loops, whereas large leakage current contributions can be seen in the positive field region, indicating a polarization direction-dependent leakage current. The polarization extracted from the saturated field matches the results we measured with PUND method. Leakage optimization of the FE films is subject of ongoing work.

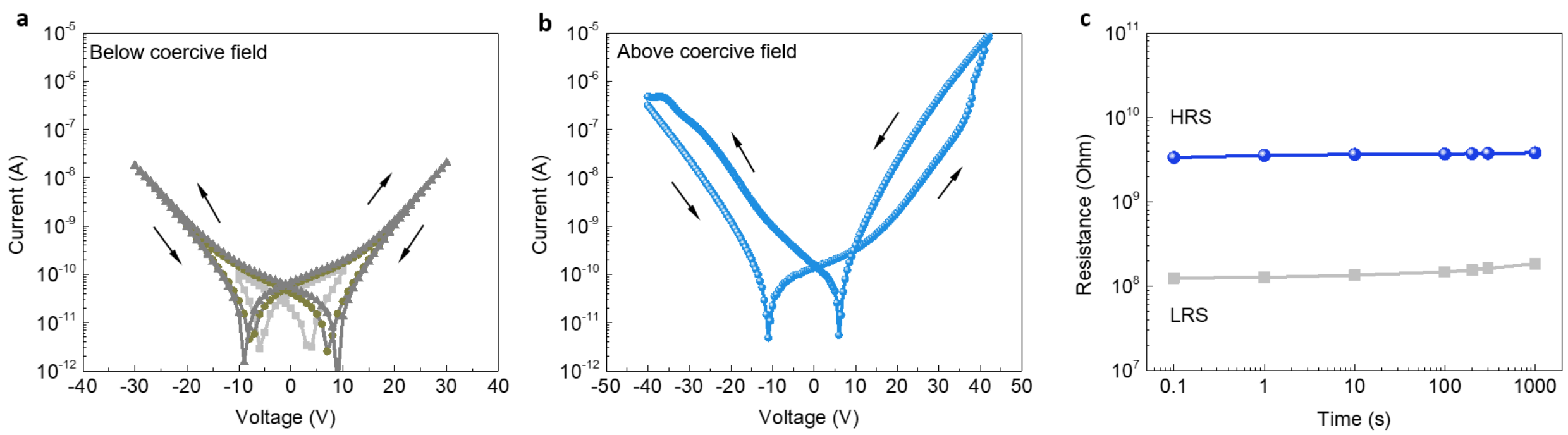

**Fig. S5. Non-volatile resistive switching of an AlScN ferroelectric MIM device diode**. a-b, Current-voltage characteristics of the demonstrated AlScN ferroelectric MIM device with applied electrical field below- and above- coercive field, respectively. Ferroelectric MIM device demonstrated non-volatile resistive switching by utilizing tunnel barrier modulation upon ferroelectric polarization switching. The current can only be switched from a high-resistance state (HRS) to a low-resistance state (LRS) by applying a voltage higher than the coercive voltages, as shown in Fig. S4a, b, which indicates the coercive voltage is between 30-40 V under DC measurement. c, Retention properties of the resistance state as a function of time obtained by programming the LRS and HRS with a gate voltage of ±40 V, then monitoring the resistance for 1,000 secs. No significant degradation was observed on both LHS and HRS.

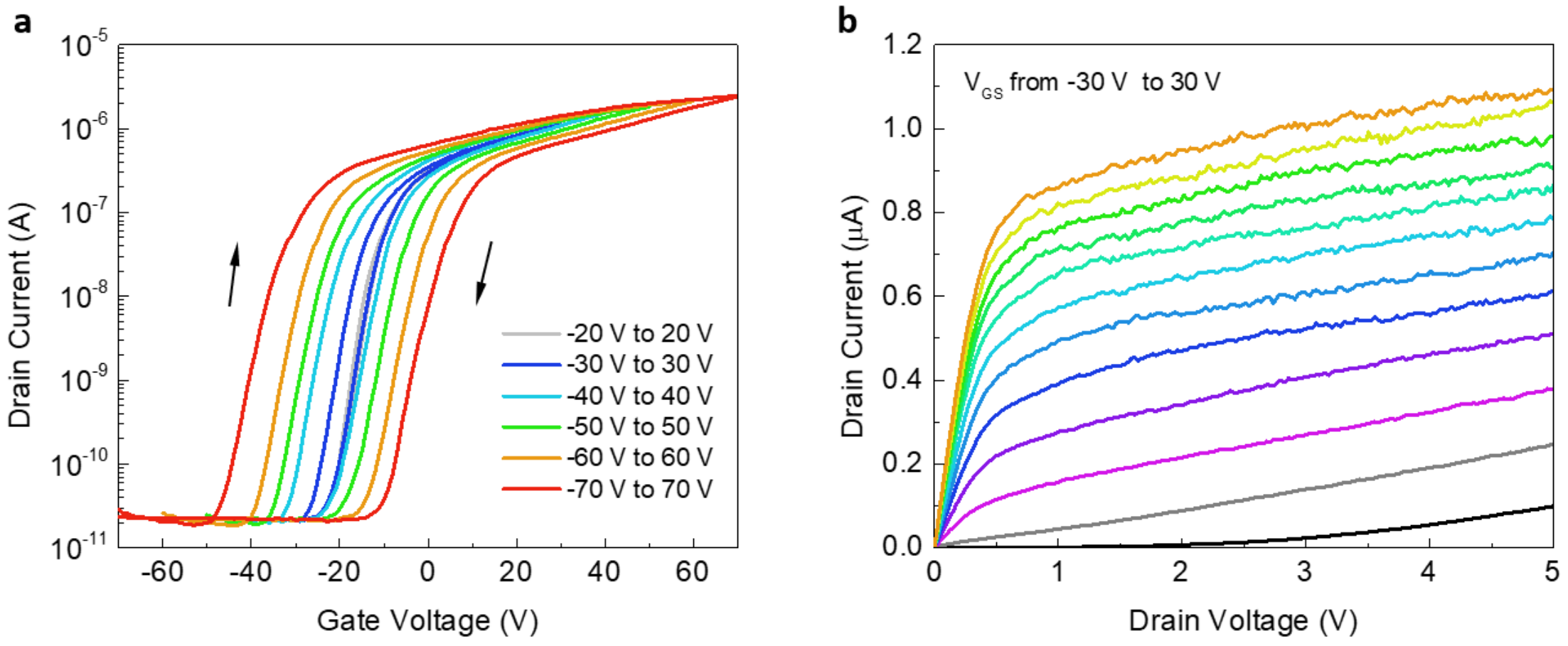

**Fig. S6. Room-temperature electrical characterization of a AlN/$MoS_2$ FET**. a, The AlN/$MoS_2$ FET was made from the same fabrication process as the AlScN/$MoS_2$ FeFETs. Source and drain electrodes are made from e-beam deposited Ti/Au (10nm/40nm). b, Linear scale output characteristics of the same device at various gate voltages showing saturation in the channel. Transfer characteristics of an AlN/$MoS_2$ FET at $V_{DS}$ = 1 V with various gate voltage sweep ranges up to ± 70 V. Significant clockwise hysteresis was observed, which is attributed to the charge trapping. The hysteresis loop direction doesn't flip sign upon a change in sweep range up to ±70 V, which serves as a strong evidence that the counterclockwise hysteresis in AlScN/$MoS_2$ FeFETs is induced by ferroelectric polarization switching.

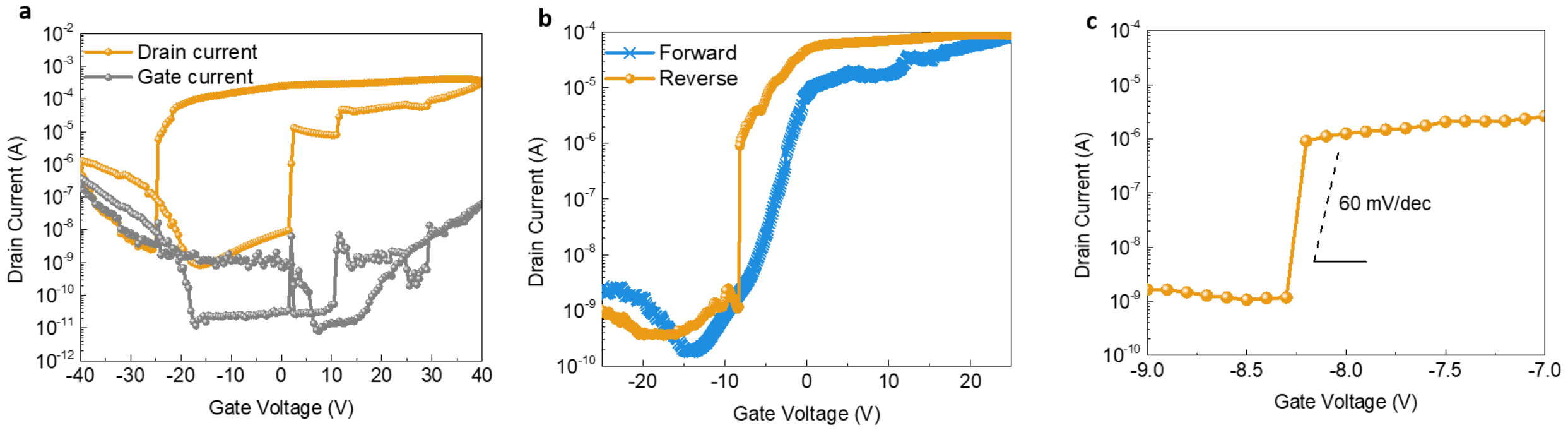

**Fig. S7. Steep sub-threshold slope induced by negative capacitance through integrating ferroelectric AlScN in the gate stack.** a, Drain- and gate- current versus gate voltage in the AlScN/$MoS_2$ FeFETs with drain volatge at 1 V. b-c, Transfer curve of AlScN/$MoS_2$ FeFETs with drain volatge at 1 V, showing a steep sub-threshold slope switching. It has been shown that a ferroelectric material integrated into the gate stack of a transistor can create an effective negative capacitance that allows the device to overcome "Boltzmann tyranny" [31-33]. The insulating ferroelectric layer serves as a negative capacitor so that the channel surface potential can be amplified more than the gate voltage, and hence the device can operate with subthreshold swing less than 60 mV/dec even at room temperatures. In a, it is interesting to notice that along with the abrupt rise or drop in drain current at about the threshold voltage, there is a simultaneous spike in the gate current that can be attributed to the amplified gate field across the gate dielectric. As shown in b and c, the subthreshold swing was observed to fall below the thermal limit ~ 60 mV/dec for over 3 decades of current, with an average of 30 mV/dec further strongly indicating FE switching in our AlScN/$MoS_2$ FET devices.

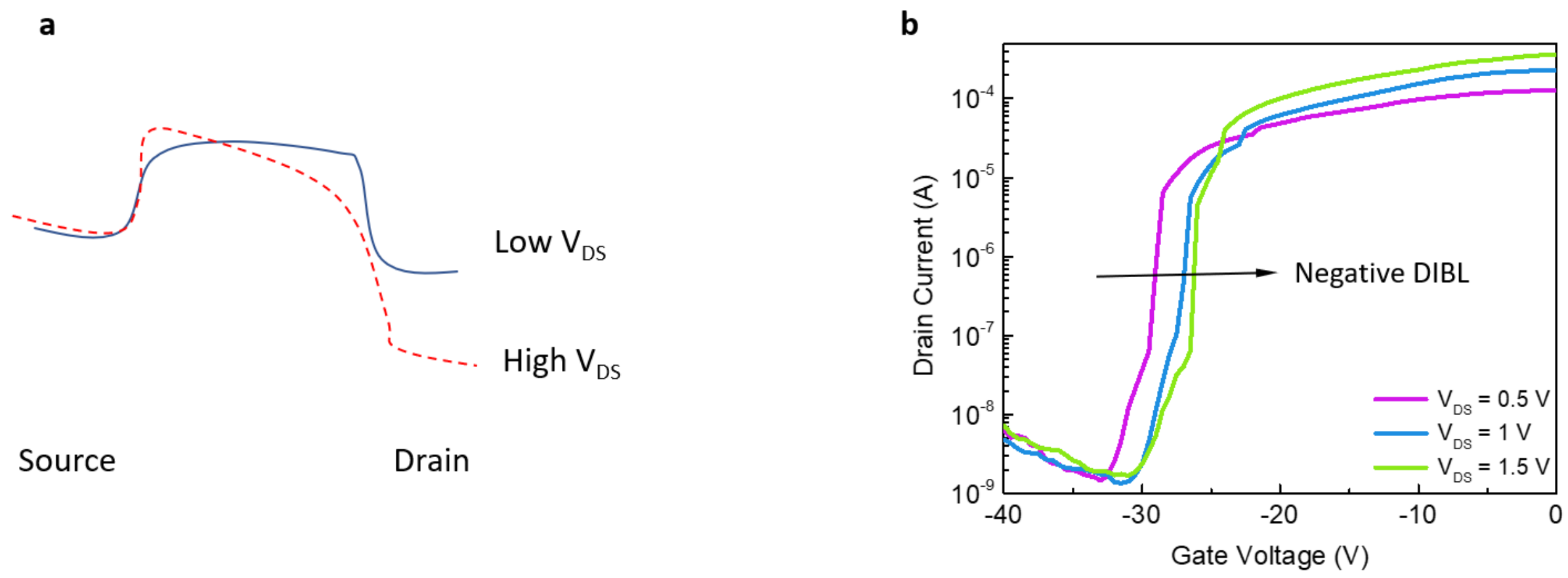

**Fig. S8. Negative drain-induced-barrier-lowering (DIBL) induced by negative capacitance in the gate stack**. a, Band diagram of the negative DIBL effect. The negative DIBL originates from capacitive coupling from the drain to the gate stack of AlScN and $MoS_2$. b, Transfer characteristics of AlScN/$MoS_2$ FETs measured at room temperature and at drain voltages of 0.5 V, 1 V and 1.5 V. A threshold voltage shift towards the positive can be observed at high drain voltage, indicating a negative DIBL effect, characteristic of an FE dielectric in the stack.

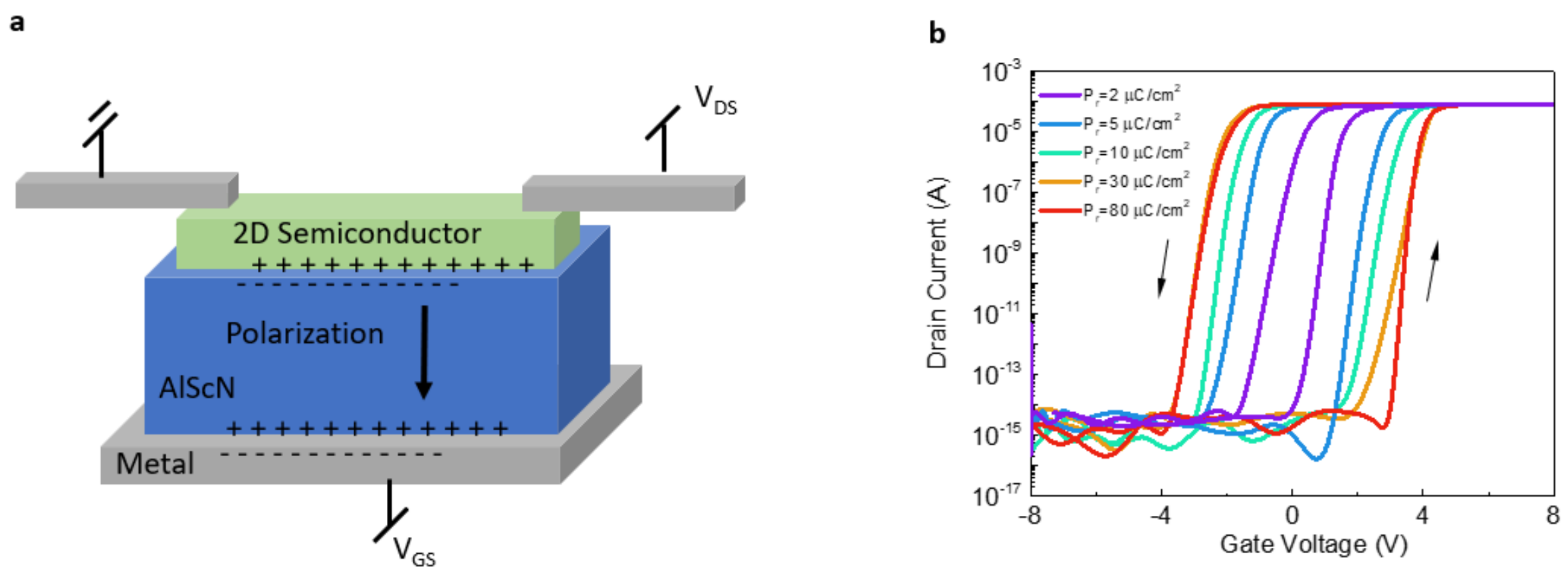

**Fig. S9. TCAD simulation of AlScN/MoS$_2$ FeFTEs with various remanent polarizations.** a, The schematic of the proposed AlScN/MoS$_2$ FeFETs in the TCAD simulation. b, Semilogarithmic scale simulated transfer characteristics at room temperature of a representative AlScN/MoS$_2$ FeFET with 100 nm AlScN as the gate ferroelectric dielectric with a channel thickness of 10 nm, a channel length of 1 μm. The AlScN/MoS$_2$ FeFET has been simulated with varying amounts of remnant polarization for the AlScN showing a large memory window that saturates for remanent polarizations > 30 μC/cm$^2$.